\documentclass[twoside]{article}
\usepackage{l3physics}   
\usepackage{fleqn}
\usepackage{tau98}
\usepackage{epsf}%
\usepackage{epsfig}%

\newcommand{\AmS}{{\protect\the\textfont2
  A\kern-.1667em\lower.5ex\hbox{M}\kern-.125emS}}

\newcommand{\errorl}{\sigma^{\mathrm{L3}}(F_2)}
\newcommand{\erroro}{\delta_{95}^{\mathrm{OPAL}}(F_2)}
\newcommand{\eeee}{{\mathrm{e^+e^-\rightarrow e^+e^-(\gamma)}}}

\renewcommand{\ee}{{\mathrm{e^+e^-}}}

\newcommand{\eemmg}{{\mathrm{e^+e^-\rightarrow\mu^+\mu^-\gamma}}}
\newcommand{\mmg}{{\mathrm{\mu\mu\gamma}}}
\newcommand{\eett}{{\mathrm{e^+e^-\rightarrow\tau^+\tau^-(\gamma)}}}
\newcommand{\eettg}{{\mathrm{e^+e^-\rightarrow\tau^+\tau^-\gamma}}}
\newcommand{\ttg}{{\mathrm{\tau\tau\gamma}}}
\newcommand{\atau}{a_\tau}
\newcommand{\dtau}{d_\tau}
\newcommand{\atausm}{a_\tau^{\scriptscriptstyle{\mathrm{SM}}}}

\renewcommand{\MZ}{m_{\scriptscriptstyle{\mathrm{Z}}}}
\newenvironment%
{myitemize}%
{\begin{list}{$\bullet$}%
{
\setlength{\itemsep}{-1.0\parsep}%
\addtolength{\itemsep}{0.6em}%
\setlength{\itemindent}{2.2ex}%
\setlength{\leftmargin}{0ex}%
\setlength{\rightmargin}{0ex}
}}%
{\end{list}}
\title{Anomalous Magnetic and Electric Dipole Moments of the Tau\\
       {\normalsize{Invited talk at the TAU'98 Workshop, 14-17 September 1998, Santander, Spain}}}

\author{Lucas Taylor\address{Department of Physics,
                             Northeastern University, Boston, MA02115, USA}}%

\begin{document}

\begin{abstract}
This paper reviews the theoretical predictions for and the 
experimental measurements of the anomalous magnetic and electric 
dipole moments of the tau lepton.
In particular, recent analyses of the $\eettg$ process
from the L3 and OPAL collaborations are described.
The most precise results, from L3, for the anomalous magnetic and 
electric dipole moments respectively are:
$\atau = 0.004 \pm 0.027 \pm 0.023$ and 
$\dtau =  (0.0  \pm 1.5 \pm 1.3)\times 10^{-16}{e{\cdot}\mathrm{cm}}$.

\end{abstract}

\maketitle

\section{INTRODUCTION}
In the Standard Model (SM), the electromagnetic interactions of each of the three 
charged leptons are assumed to be identical, apart from mass effects.
There is, however, no experimentally verified explanation for why there are three generations 
of leptons nor for why they have such differing masses.
New insight might be forthcoming if the leptons were observed to have substructure
which could manifest itself experimentally in anomalous values for the magnetic or 
electric dipole moments.

In general a photon may couple to a lepton through its electric charge, 
magnetic dipole moment, or electric dipole moment 
(neglecting possible anapole moments\cite{MARCIANO90A}).
The most general Lorentz-invariant form for the coupling of a charged lepton to a photon 
of 4-momentum $q_\nu$ is obtained by replacing the usual $\gamma^\mu$ by
\begin{eqnarray} 
 \Gamma^\mu & = & F_1(q^2) \gamma^\mu                                  \nonumber \\
            & + & F_2(q^2) \frac{i}{2m_\ell}\sigma^{\mu\nu}q_\nu        
              -   F_3(q^2) \sigma^{\mu\nu}\gamma^5 q_\nu. 
\label{equ:gamma}
\end{eqnarray} 
The $q^2$-dependent form-factors, $F_i(q^2)$, have familiar interpretations for $q^2=0$.
$F_1(q^2\!=\!0) \equiv Q_\ell$ is the electric charge of the lepton. 
$F_2(q^2\!=\!0) \equiv a_\ell=(g_\ell-2)/2$ is the anomalous magnetic moment of the lepton.
$F_3(q^2\!=\!0) \equiv d_\ell/Q_\ell$, where $d_\ell$ is the electric dipole moment of the lepton.
Hermiticity of the electromagnetic current forces all of the $F_i$ to be real.
In the SM $a_\ell$ is non-zero due to loop diagrams.
A non-zero value of $d_\ell$ is forbidden by both $P$ invariance and $T$ invariance 
such that, if $CPT$ invariance is assumed, observation of a non-zero value of $d_\ell$ would 
imply $CP$ violation.

The anomalous moments for the electron and muon have been 
measured with very high precision and are in impressive 
agreement with the theoretical predictions, as shown in 
Table~\ref{tab:eandmu}.
\begin{table}[!htb]
\setlength{\tabcolsep}{1.0pc}
\caption{Theoretical predictions and experimental measurements of the
         anomalous magnetic and electric dipole moments of the electron and muon.}
\label{tab:eandmu}
\begin{tabular*}{0.47\textwidth}{llc}  \hline  
$a_{\mathrm{e}}^{\mathrm{\scriptscriptstyle{SM}}}$ & $=0.001\,159\,652\,46(15)$  & \cite{KINOSHITA81A} \\
$a_{\mathrm{e}}^{\mathrm{expt}}$                   & $=0.001\,159\,652\,193(10)$ & \cite{COHEN87A}     \\  \hline   
$a_\mu^{\mathrm{\scriptscriptstyle{SM}}}$          & $=0.001\,165\,920\,2(20)$   & \cite{HUGHES85A}    \\
$a_\mu^{\mathrm{expt}}$                            & $=0.001\,165\,923\,0(84)$   & \cite{COHEN87A}     \\ \hline  
$d_{\mathrm{e}}^{\mathrm{expt}}$& $=(-2.7 \pm 8.3) \cdot 10^{-27}\,e\,{\mathrm{cm}}$  & \cite{ABDULLAH90A}   \\  \hline  
$d_\mu^{\mathrm{expt}}$         & $=( 3.7 \pm 3.4) \cdot 10^{-19}\,e\,{\mathrm{cm}}$  & \cite{BAILEY78A}     \\  \hline  
\end{tabular*}
\end{table}

In the SM $\atau$ is predicted to be
$\atausm = 0.001\,177\,3(3)$~\cite{SAMUEL91A,HAMZEH96A}.
The short lifetime of the tau precludes precession measurements 
which means that the tau cannot be measured with a precision 
comparable to those achieved for the electron and muon.
Less precise experimental measurements of $\atau$ and $\dtau$ 
using different techniques are nonetheless extremely interesting 
since they are sensitive to a wide variety of new physics 
processes.

The process 
$\mathrm{e}^+\mathrm{e}^- \rightarrow \gamma \rightarrow \tau^+\tau^-$
has been used to constrain $F_2$ and $F_3$ of the tau at $q^2$ up to
$(37~\mathrm{GeV})^2$~\cite{SILVERMAN83A}, and an indirect limit has been inferred
from the width $\Gamma(Z \rightarrow \tau^+ \tau^-)$~\cite{ESCRIBANO97A}.
These results do not correspond to $q^2=0$ therefore the upper limits obtained
on the form factors $F_2$ and $F_3$ do not correspond to constraints on 
the static tau properties, $a_\tau$ and $d_\tau$.

The process $\eettg$, first analysed by Grifols and Mendez 
using L3 data~\cite{GRIFOLS91A}, is attractive since $q^2 = 0$ for the on-shell 
photon\footnote{Strictly speaking, the $F_i$ are functions of 
                three variables, $F_i(q^2,m_1^2,m_2^2)$, where $m_1$ and 
                $m_2$ are the $\tau$ masses on either side of the $\tau\tau\gamma$ 
                vertex, such that $q^2=0$ and $m_1^2=m_\tau^2$ but $m_2^2$ 
                corresponds to an off-shell $\tau$.}.
In this paper, we report on recent measurements of this process  
from the L3 and OPAL collaborations~\cite{L3TTG,OPALTTG}.
These are based on analyses of $\eettg$ events selected from the full LEP 
I on-peak data samples.
No results are currently available from the ALEPH, DELPHI, or 
SLD collaborations.

\section{THEORETICAL MODELLING}
There are two published calculations of the process $\eettg$, allowing for 
anomalous electromagnetic moments,
that of Biebel and Riemann, which we refer to as ``B\&R''~\cite{BIEBEL96A},
and that of Gau, Paul, Swain, and Taylor which is known 
as ``TTG''~\cite{TTGNUCPHYSB,PAULWASPAPER}.

L3 uses the TTG calculation while OPAL uses an unpublished calculation 
of D. Zeppenfeld, with an additional correction based on the 
B\&R calculation.

\subsection{Comparison of theoretical models}

\begin{figure}[!tb]
\begin{center}
\epsfig{file=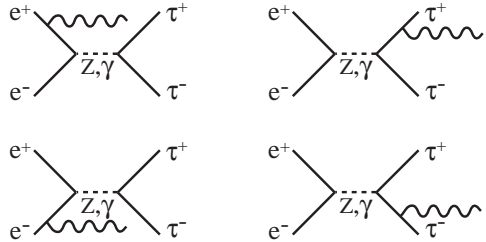,width=0.4\textwidth,clip=}
\caption{\label{fig:feynmann}Diagrams contributing to $e^+ e^- 
\rightarrow \tau^+ \tau^- \gamma$ at tree-level.}
\end{center}
\end{figure}

Both B\&R and TTG parametrise the effects of anomalous dipole moments using 
the ansatz of Eq.~\ref{equ:gamma} and calculate the 
squared matrix element for the process $\eettg$.
The TTG calculation is more complete than that of ``B\&R'' and 
includes all the Standard Model and anomalous amplitudes for the 
diagrams shown in Figure~\ref{fig:feynmann}.  

The TTG matrix element, ${\cal{M}}(\atau,\dtau)$,  is determined without making 
simplifying assumptions. 
In particular, no interference terms are neglected and no fermion masses 
are assumed to be zero.  
The inclusion of a non-zero tau mass is essential, as
the (significant) interference terms between the Standard Model and
the anomalous  amplitudes vanish in the limit of vanishing tau mass. 
Standard Model radiative corrections are incorporated by
using the improved Born approximation.

Infrared divergences for collinear or low energy photons may be
naturally avoided by applying cuts on minimum photon energy and/or
opening angle between the photon and tau.  
This method eliminates potential concerns about the consistency of 
cancelling divergences against vertex corrections which are calculated 
in QED assuming that the anomalous couplings vanish. 
Moreover, this procedure matches well the experimental reality, since 
electromagnetic calorimeters have a minimum energy cutoff and isolation cuts 
are required to distinguish tau decay products from radiated photons.

The TTG calculation of ${\cal{M}}(0,0)$ is checked by 
comparison with the SM $O(\alpha)$ predictions of KORALZ.  
The shapes of the photon energy and angular distributions are in 
excellent agreement and the overall normalisation agrees to $0.1\%$. 

The B\&R calculation makes some reasonable approximations in order to 
arrive at a more manageable analytic expression, compared to the 
very complicated one provided by TTG.
In particular, B\&R neglects anomalous contributions to the ISR and FSR 
interference, $\gamma$ exchange, and $\gamma-$Z interference.
To check the two calculations, the quantity
$\Delta{\cal{M}}^2 = |{\cal{M}}(\atau,\dtau)|^2 - |{\cal{M}}(0,0)|^2$ from 
TTG is compared to the prediction of B\&R, using the same approximations in TTG
as used in the B\&R calculation.
The anomalous contribution to the cross section agrees to better than 
$1\%$~\cite{TTGNUCPHYSB} and the shapes of the photon energy spectra are in 
good agreement for a wide range of $\atau$ and $\dtau$ values.
The B\&R calculation differs from the {\em{full}} TTG calculation by 
only 1\% indicating that the neglect of certain terms in the B\&R case
is valid within the sensitivity of the LEP experiments. 

\begin{figure}[!tb]
\begin{center}
\epsfig{file=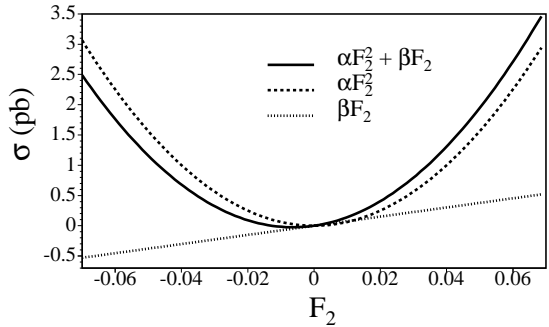,width=0.45\textwidth,clip=} 
\caption{\label{fig:x_mdm}Anomalous contribution to the
$e^+e^-\rightarrow\tau^+\tau^-\gamma$ cross section as a function
of $F_2(0)$ showing the linear and quadratic contributions.}
\end{center}
\end{figure}

Both calculations show that terms linear in $F_2(0)$  
arise from interference between Standard Model and anomalous amplitudes. 
Figure~\ref{fig:x_mdm} shows the
contribution to the total cross section arising from these terms, with
the linear and quadratic components shown separately~\cite{TTGNUCPHYSB}. 
The interference terms for $F_2(0)$ are significant compared to the total 
anomalous cross section for small values of $F_2(0)$. 
For example, for $F_2(0)=0.01$, inclusion of the linear terms enhances the number of 
excess photons, compared to the purely quadratic calculation, by 
a factor of approximately five.
The anomalous contribution due to $F_3(0)$ is identical to the 
quadratic term for $F_2(0)$ in that the linear terms vanish identically.
%

\subsection{Effects of anomalous couplings}

\begin{figure*}[!tb]
\begin{center}
\epsfig{file=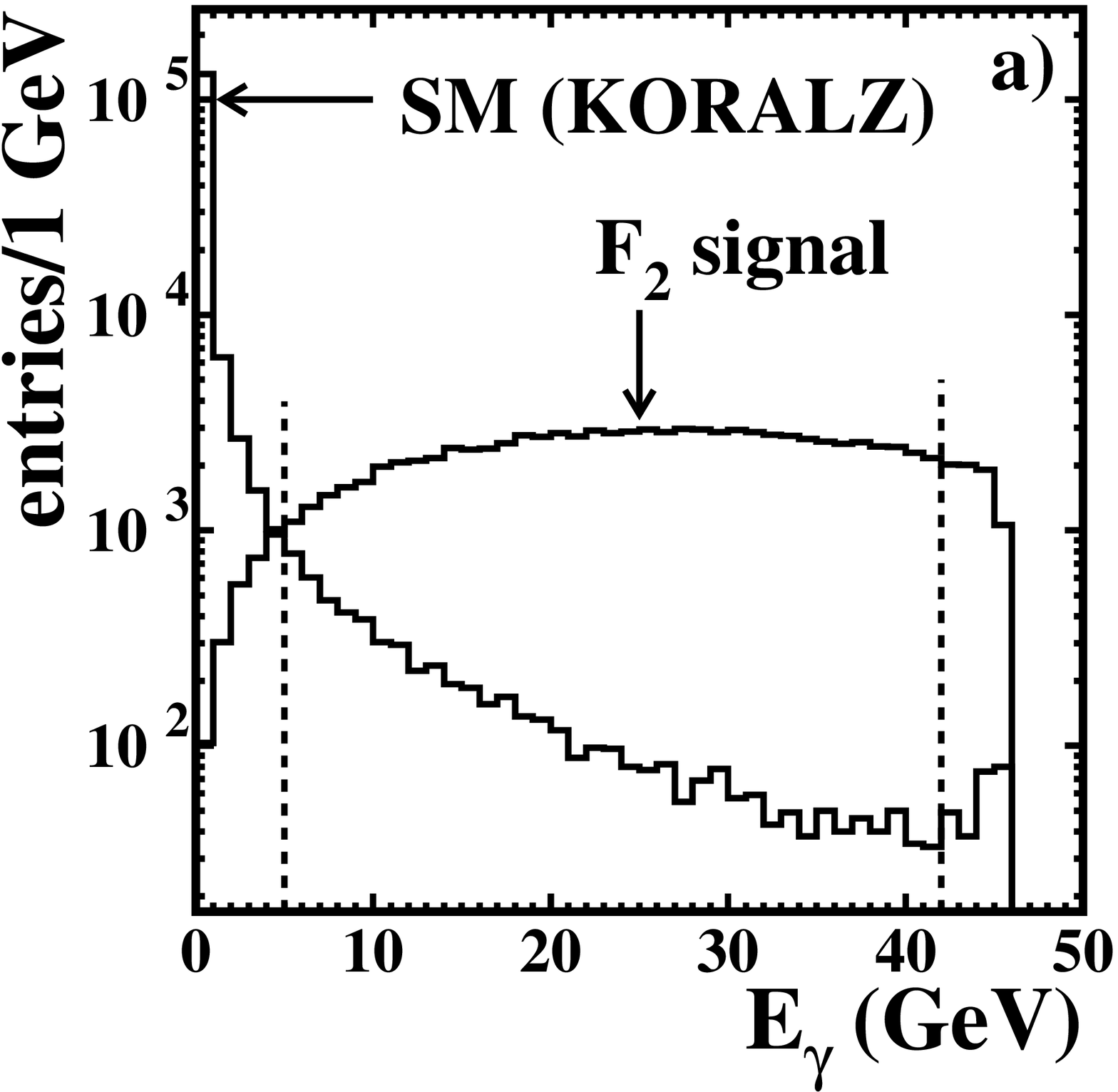,width=0.32\textwidth,clip=}~~%
\epsfig{file=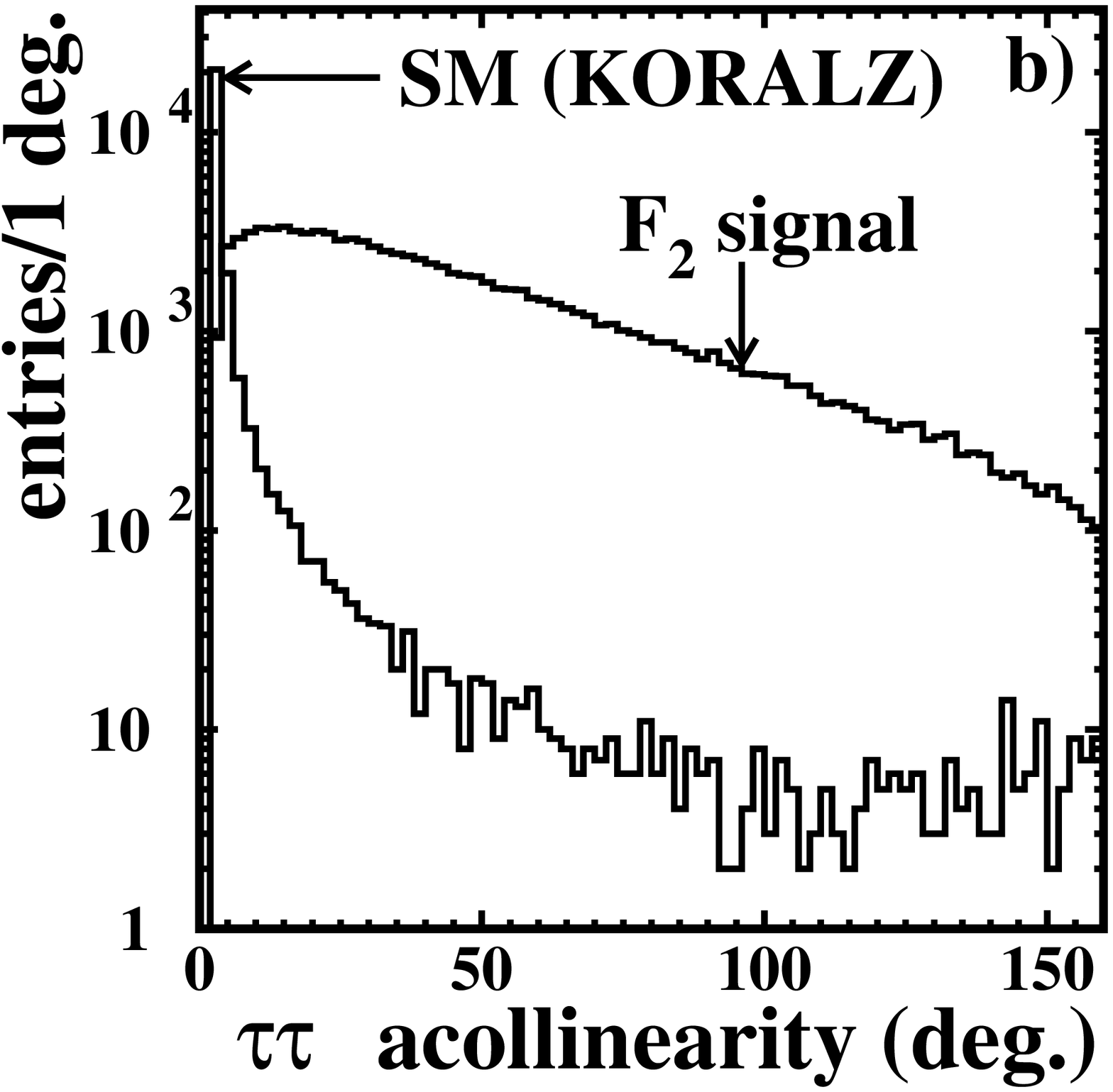,width=0.32\textwidth,clip=}~~%
\epsfig{file=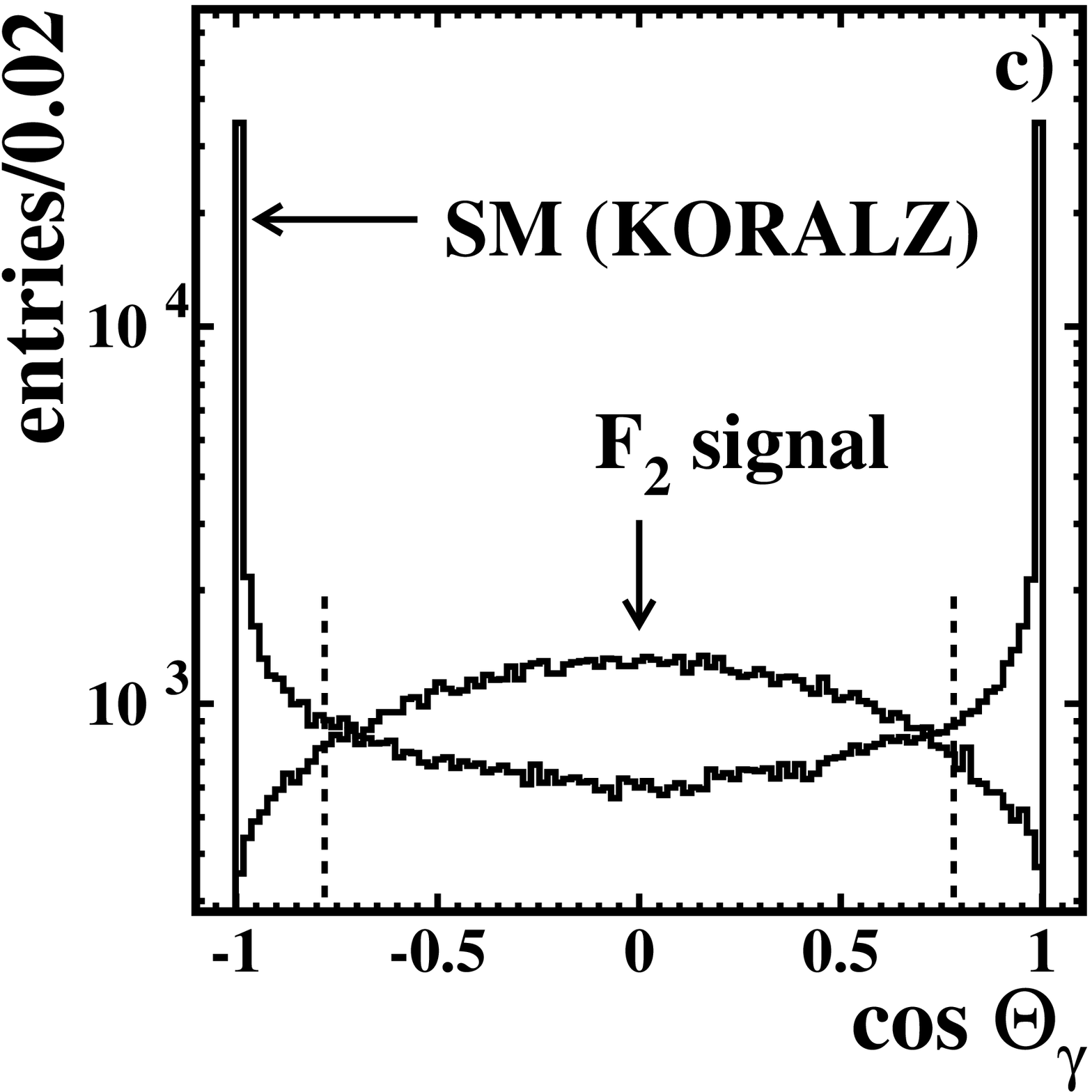,width=0.32\textwidth,clip=}%
\caption{\label{fig:opal_theory}%
OPAL Monte Carlo study of 
a) the photon energy spectrum, 
b) the $\tau\tau$ acollinearity, and 
c) the polar angle of the photon in the laboratory frame 
for the SM (KORALZ) contribution and the arbitrarily-normalised 
anomalous contribution from a non-zero $F_2(0)$.}
\end{center}
\end{figure*}

Figure~\ref{fig:opal_theory} shows the results of an OPAL Monte Carlo 
study of the photon energy spectrum, $\tau\tau$ acollinearity, and 
the polar angle of the photon in the laboratory frame for the 
SM, represented by the KORALZ Monte Carlo~\cite{KORALZ}, 
and the shape of the additional anomalous contribution.

The anomalous electromagnetic moments tend to enhance the production of 
high energy photons which are isolated from the taus, compared to the 
SM final state radiation which has a rapidly falling energy 
spectrum of photons which tend to be collinear with one of the taus.
The characteristics of the energy and angular distributions of $\eettg$ events 
are exploited to enhance the sensitivity to 
anomalous moments, as described below.

\subsection{Monte Carlo events samples}

In order to study the various selection criteria and the backgrounds
L3 and OPAL use large samples of Monte Carlo events,
in particular:
hadronic            events from the JETSET Monte Carlo program~\cite{jetset};
two-photon          events from DIAG36~\cite{DIAG36} (L3) and VERMASEREN (OPAL)~\cite[ref. 17]{OPALTTG};
Bhabha              events from BHAGENE~\cite{bhagenenew} (L3) or RADBAB (OPAL)~\cite[ref. 17]{OPALTTG}; 
$\mu\mu(\gamma)$    and 
$\tau\tau(\gamma)$  events from KORALZ~\cite{KORALZLONG}.
The KORALZ~ samples include the effects of initial and final state SM
bremsstrahlung corrections to $O(\alpha ^2)$ including exclusive exponentiation.
All Monte Carlo events are passed through GEANT-based detector 
simulation programs, and reconstructed in the same 
way the data.

To model the effects of anomalous moments allowing for all SM, detector, and 
reconstruction effects the KORALZ samples of SM $\eettg$ events are reweighted 
according to various assumed values of $F_2(0)$ and $F_3(0)$.

In the L3 case, TTG is used to determine a weight for each KORALZ event which 
depends on the generated four-vectors of the taus and the photon and on the 
values of $F_2(0)$ and $F_3(0)$ under consideration.
This procedure takes full account of any dependences of the 
acceptance or reconstruction efficiency which arise from the 
kinematic variations of the event topologies with $F_2(0)$ and $F_3(0)$.
Since TTG is an $O(\alpha)$ calculation, there is no unambiguous way to 
compute a weight for events with more than one photon.  
The treatment of multiple photon events is treated as a systematic error.

In the OPAL case, individual events are not reweighted. 
Instead, only the selected photon energy distribution is reweighted according to 
the naive Zeppenfeld predictions (which neglect interference completely)
and a correction is then applied for the missing interference terms 
using the B\&R prediction.
            
\section{EXPERIMENTAL ANALYSIS}
L3 and OPAL both analyse their full on-peak Z data samples which each correspond 
to an integrated luminosity of approximately 100\,${\mathrm{pb}}^{-1}$.

\subsection{Event selection}

L3 and OPAL first select a sample enriched in $\eett$ events by rejecting 
most of the hadronic, Bhabha, dimuon, two-photon, and other background 
events using reasonably standard selection cuts.
Then, they identify photons above a few GeV in energy which are isolated 
from the decay products of the tau. 
            
\begin{figure}[htb!]
\begin{center}
\framebox[0.43\textwidth]{\epsfig{file=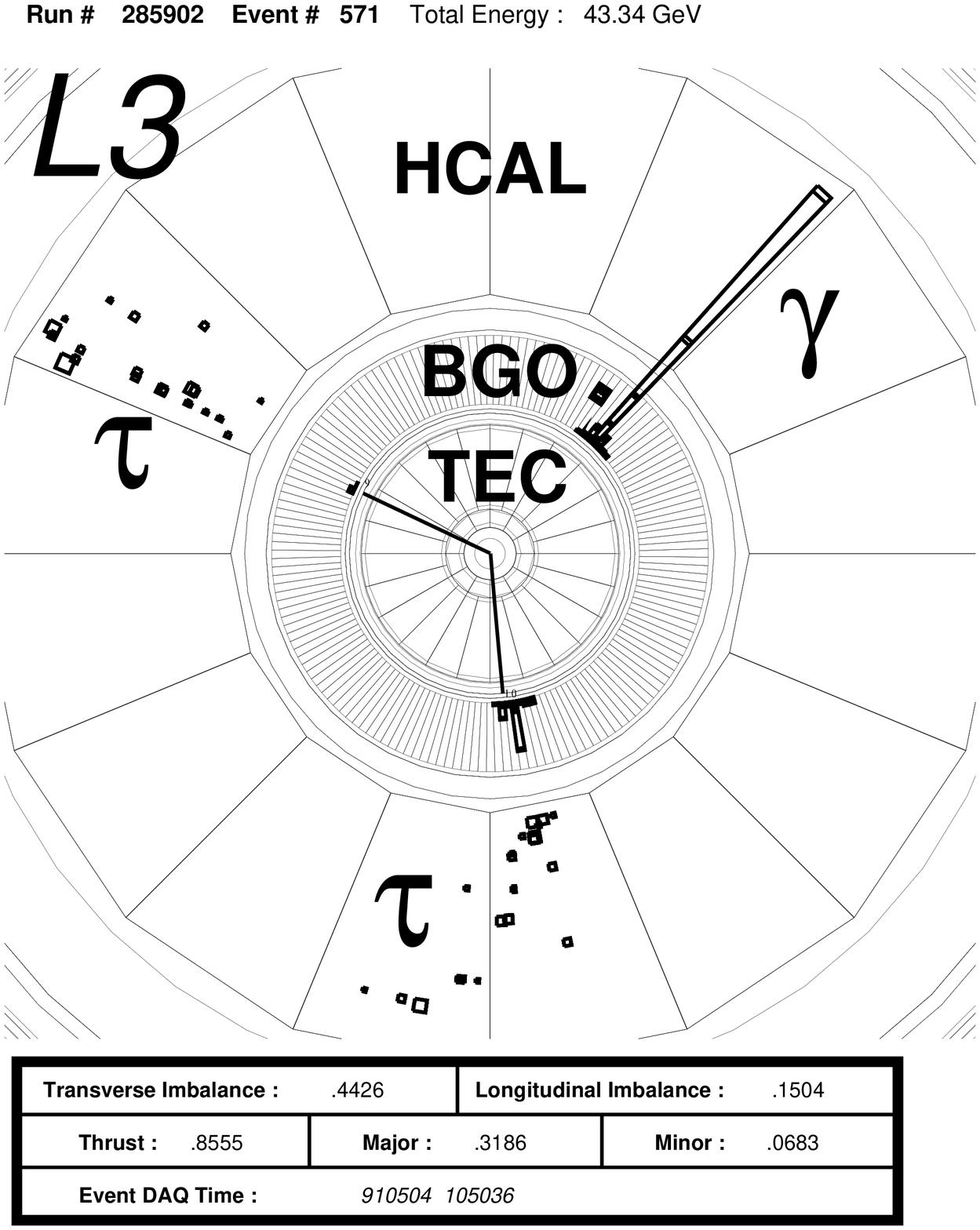,width=0.4\textwidth,clip=}}
\end{center}
\caption{Typical $\tau\tau\gamma$ candidate event at L3.}
\label{fig:event}
\end{figure}

These events are in general very distinctive.
For example, Figure~\ref{fig:event} shows a typical $\tau\tau\gamma$ candidate 
event as seen in the L3 detector.
The decay products of the taus in this event leave tracks in the inner tracker (TEC)
and energy in the electromagetic (BGO) and hadron calorimeters.
The photon has no associated track, leaves a considerable amount of 
energy in the BGO, and no energy in the hadron calorimeter.

L3 selects 1559 $\eettg$ candidate events with a non-$\ttg$ background 
of 6.7\% while OPAL selects 1429 events with a background of 0.13\%.
The Monte Carlo samples show that for both experiments the $\ttg$ samples are 
dominated by genuine SM $\ttg$ events.
For example, SM processes are predicted to yield a sample of 1590 events 
in the L3 case.
The significant difference in non-$\ttg$ background levels reflects 
differing severities of selection cuts.
For example, OPAL uses electron momentum measurements in the tracker to 
reject residual Bhabha backgrounds, which is not done for the L3 analysis. 

\subsection{Determination of $F_2(0)$ and $F_3(0)$}
Anomalous values of $F_2(0)$ and $F_3(0)$ tend to increase the cross section
for the process $\eettg$, especially for photons with high energy which are
well isolated from the decay products of the taus.  
The information used in the fits to extract $F_2(0)$ and $F_3(0)$ are somewhat 
different for the two experiments as described below, although 
both L3 and OPAL conservatively set $F_3(0) = 0$ when fitting for $F_2(0)$ 
and {\it vice versa}. 

\begin{figure}[!tb]
\begin{center}
     \mbox{\epsfig{file=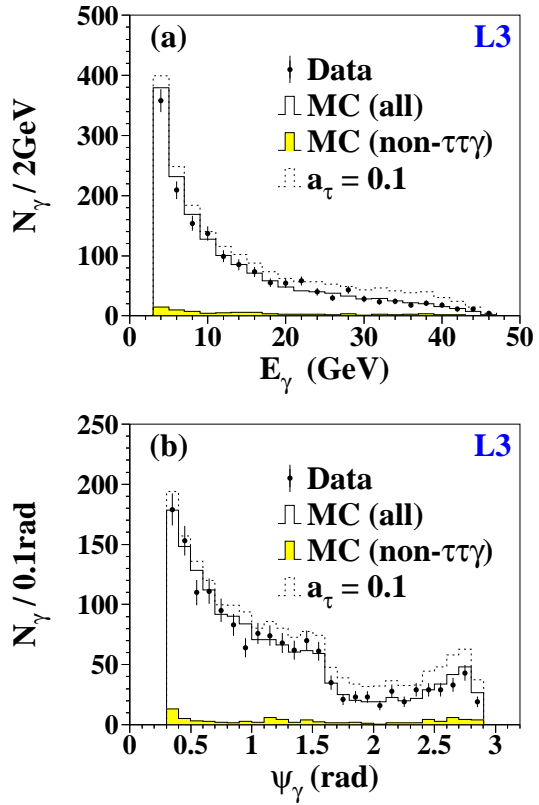,width=0.45\textwidth,clip=}} 
\caption{The number $N_\gamma$ of photon candidates in the L3 $\eettg$ sample
         as a function of 
         (a) $E_\gamma$ and 
         (b) $\psi_\gamma$.         
         The points with error bars denote the data and the 
         solid histograms denote the SM Monte Carlo predictions.  
         The dashed histograms show how the predictions for 
         $\atau = 0.1$.  
\label{fig:gameandisol}} 
\end{center}
\end{figure}

L3 makes binned maximum likelihood fits to the two-dimensional distribution of 
the photon energy, $E_\gamma$, {\em{vs.}} the angle between the photon and 
the closest tau jet, $\psi_\gamma$. 
To exploit the cross-section information, the SM Monte Carlo samples are 
normalised to the integrated luminosity.
Figure~\ref{fig:gameandisol} shows the L3 distributions of $E_\gamma$ and 
$\psi_\gamma$ for the data and the SM Monte Carlo expectation.
Both the increase in the total cross section and the relatively greater 
importance of photons with large $E_\gamma$ and $\psi_\gamma$
are evident.          
No excess is apparent at high values as would be expected for significant
deviations of $F_2(0)$ and/or $F_3(0)$ from their SM values.
The results of the L3 fits to the data, considering only statistical errors, are 
$F_2(0) = 0.004 \pm 0.027$ and  
$F_3(0) = (0.0 \pm  1.5) \times 10^{-16} e \cdot \mathrm{cm}$,
where the errors refer to the $68.3\%$ confidence interval.
These two results are not independent, although the absence of interference terms 
for $\dtau$ does provide some distinguishing power between the effects of 
$\atau$ and those of $\dtau$~\cite{TTGNUCPHYSB}.

\begin{figure}[!tb]
\begin{center}
     \mbox{\epsfig{file=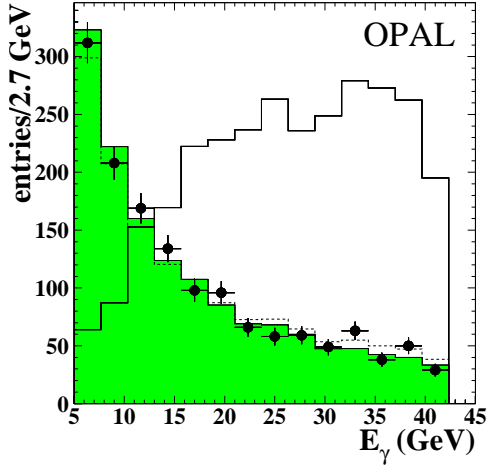,width=0.4\textwidth,clip=}} 
\caption{The number of photon candidates in the OPAL $\eettg$ sample
         as a function of $E_\gamma$.         
         The points with error bars denote the data and the filled
         solid-line histogram denotes the SM prediction.  
         The dashed histogram shows the expectation for $F_2(0) = 0.064$.  
         The un-filled solid-line histogram shows, with arbitrary 
         normalisation, the signal expectation only.           
\label{fig:opalenergy}} 
\end{center}
\end{figure}

OPAL makes a one-dimensional fit to the energy spectrum of the photons.
The total cross-section information is not used since the normalisation 
of data and Monte Carlo are forced to agree.
Although there is some loss of sensitivity as a result, there is a 
negligible systematic error from uncertainties in the photon
reconstruction efficiency.
Figure~\ref{fig:opalenergy} shows the OPAL distributions of photon 
energy for the data and the SM Monte Carlo expectation.
The OPAL result for $F_2(0)$, after correction for the effects of 
interference, is $0.065 < F_2(0) < 0.062$ at the 95\% confidence level.
The value of $F_3(0)$ is not fitted directly but is inferred from 
the $F_2(0)$ analysis, as described below.

\subsection{Systematic errors}

Both L3 and OPAL perform numerous cross-checks of their analyses using 
independent data samples and the Monte Carlo samples.
The various contributions to the systematic errors on $F_2(0)$ 
are described below, those for $F_3(0)$ are similar.

It is important to note that while L3 quotes individual systematic errors
in the conventional ($\pm 1\sigma$) way, the OPAL values correspond to
estimated changes in the 95\% limit and are hence numerically much 
smaller in general.
To make this distinction clear, we denote the former by 
$\errorl$ and the latter by $\erroro$.

\begin{myitemize}
\item {\bf{Event selection cuts}}\\
Wide variations in the selection cuts yield systematics of  
$\errorl = 0.013$ and $\erroro = 0.005$.

\item {\bf{$\tau\tau(\gamma)$ selection efficiency}}\\
To verify the $\eett$ event selection efficiency, L3 determines the $\eett$ 
cross-section at $\sqrt{s}\approx m_{\mathrm{Z}}$ to be
$\sigma_{\tau\tau} = (1.472 \pm 0.006 \pm 0.020)\,{\rm{nb}}$,
in agreement with the SM prediction of ZFITTER~\cite{ZFITTERLTHREE} of 
$1.479{\rm{\,nb}}$.
This indicates the absence of significant systematic effects 
in the selection of taus and contributes an error of $\errorl = 0.011$.

OPAL is relatively insensitive to this source of uncertainty since the 
normalisation of the data and Monte Carlo is enforced 
(resulting in a loss of statistical sensitivity).

\item {\bf{Photon reconstruction efficiency}}\\
L3 performed a particularly compelling analysis of $\eemmg$ events 
selected from the data and compared to Monte Carlo predictions,
which are known to have no significant anomalous effects.
Figure~\ref{fig:mmg} shows the distributions of photon energy $E_\gamma$  
and the isolation angle $\psi_{\mu \gamma}$ of the photon to the closest muon in 
the selected $\mu\mu\gamma$ event sample.
The data are in good agreement with the Monte Carlo prediction.
The ratio of the number of photons in data to the number in the 
Monte Carlo is $0.993 \pm 0.013 \pm 0.003$.
The shapes of the energy and isolation distributions also agree well,
with $\chi^2/NDF = 50.2 / 44$ for the former and 
$\chi^2/NDF = 37.6 / 30$ for the latter, based only on the statistical error.
\begin{figure}[!tb]
\begin{center}
      \mbox{\epsfig{file=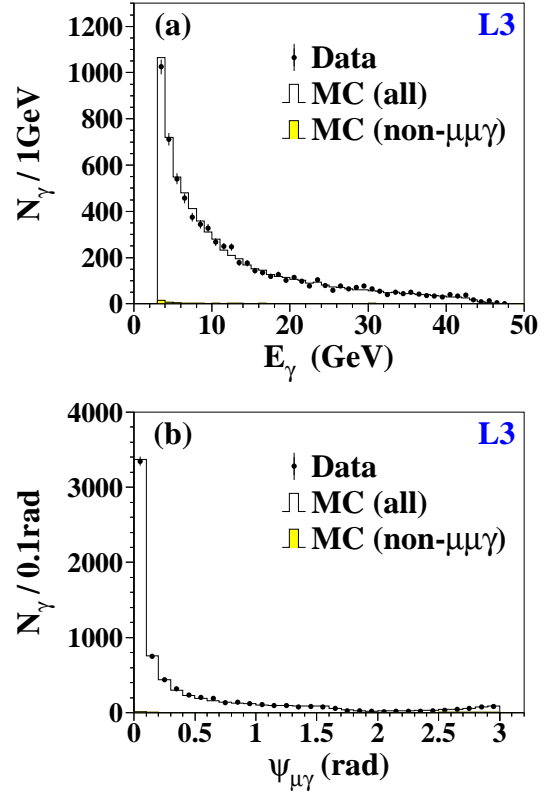,width=0.45\textwidth,clip=}} 
\caption{The number $N_\gamma$ of photon candidates in the $\eemmg$ sample
         as a function of 
         (a) $E_\gamma$ and 
         (b) $\psi_{\mu\gamma}$.         
         The points with error bars denote the data and the 
         histograms denote the Monte Carlo predictions.
\label{fig:mmg}} 
\end{center}
\end{figure}
The uncertainty on the photon reconstruction efficiency yields 
an error of $\errorl = 0.010$, which is conservative since some effects 
are covered by the variation of the selection cuts.

OPAL is relatively insensitive to this source of uncertainty since the 
normalisation of the data and Monte Carlo is enforced.

\item {\bf{Backgrounds}}\\
The L3 error due to uncertainties in the non-$\ttg$ 
backgrounds is $\errorl = 0.009$.
By comparison, OPAL has negligible uncertainty due to their  
lower level of non-$\ttg$ background.

\item {\bf{Binning}}\\
The selected samples are fitted using a variety of binning schemes, 
to yield $\errorl = 0.008$ for the two-dimensional L3 fit 
and $\erroro = 0.002$ for the one-dimensional OPAL fit.

\item {\bf{Photon energy scale and resolution}}\\
L3 studies electrons in $\eeee$, $\eett$, and ${\mathrm{e^+e^-e^+e^-}}$ events,
pairs of photons from $\pi^0$ decays, and pairs of electrons from $J/\psi$ decays.
The photon energy scale uncertainty is 
$<$0.5\%  for $E_\gamma\approx 3$\,GeV and 
$<$0.05\% for $E_\gamma\approx \MZ/2$.
The resolution is 
$(1.7 \pm 0.3\%)$ at $E_\gamma\approx 10$\,GeV and 
$(1.4 \pm 0.1)\%$ for $E_\gamma\approx \MZ/2$. 
Allowing for theses uncertainties yields $\errorl = 0.008$.
The $\mmg$ study also confirms the absence of significant systematic uncertainties in
the photon energy measurement.

OPAL verifies their energy scale somewhat less precisely (0.9\%)
using $\pi^0$ decays and estimates an effect of $\erroro = 0.001$.
They do not quote a systematic error from uncertainties in 
the photon energy resolution.

\item {\bf{Modelling of $\eettg$}}\\
The inclusion by L3 of the $\mmg$ error allows for possible systematics 
in the KORALZ description of SM photon radiation.
The TTG calculation of ${\cal{M}}(0,0)$, used by L3, agrees with the 
$O(\alpha)$ predictions of KORALZ to within $0.1\%$. 
The TTG calculation of $|{\cal{M}}(\atau,\dtau)|^2$ agrees, for the same 
approximations, with the B\&R calculation~\cite{BIEBEL96A}
to better than $1\%$~\cite{TTGNUCPHYSB}.
Variation of the TTG predictions within these uncertainties causes a 
negligible change in the L3 fit results.  

OPAL does not explicitly comment on this potential source 
of systematic uncertainty.

\item {\bf{Multiple photon radiation}}\\
To estimate the effects of multiple photon radiation in the L3 analysis,
KORALZ is used to generate 
a sample of ${\mathrm{e^+e^-}} \rightarrow \tau^+\tau^-(n\gamma)$ events.
Then all photons, except for the one with the highest momentum transverse to 
the closer tau, are incorporated into the four-vectors of the other particles  
in such a way that all particles remain on mass shell~\cite{PAULWASPAPER}.
Weights for various $\atau$ and $\dtau$ are then computed by TTG using the
modified four-vectors of the taus and the photon.  
Taking these weights in lieu of those computed using the previously described 
method, in which only events with a single hard photon are considered, has 
a negligible effect on the result of the fit.

OPAL does not explicitly comment on this potential source 
of systematic uncertainty.

\end{myitemize}

\section{SUMMARY}
The systematic errors described above are combined, to yield
the L3 fit results for $\atau$ and $\dtau$:
\begin{eqnarray}
\atau & = & 0.004 \pm 0.027 \pm 0.023 \\  
\dtau & = & (0.0 \pm 1.5 \pm 1.3) \times 10^{-16} e \cdot \mathrm{cm} 
\end{eqnarray}
where the first error is statistical and the second error is systematic.
These correspond to the following limits:
\begin{equation}
-0.052  < \atau < 0.058~~~{\mathrm{and}}
\end{equation}
\begin{equation}
|\dtau| < 3.1 \times 10^{-16}e\cdot\mathrm{cm}
\end{equation}
at the 95\% confidence level.  

OPAL chooses to quote 95\% confidence level limits only.
Their result is 
\begin{equation}
-0.068 < \atau < 0.065. 
\end{equation}
If interference is neglected, the calculations for the magnetic and electric 
dipole moments are equivalent.
Therefore, OPAL makes the substitution:
 $F_2 / 2 m_\tau \longrightarrow F_3$. 
to convert their constraints on $|F_2|$ to a constraint 
on $F_3$: 
\begin{equation}
|\dtau| < 3.7 \times 10^{-16}e\cdot\mathrm{cm}.
\end{equation}

The OPAL results are comparable with the slightly more precise L3 results.
Both are consistent with the SM expectations and show no evidence for 
the effects of new physics.

\section*{ACKNOWLEDGEMENTS}
I would like to thank Tom Paul and John Swain of L3 and Norbert Wermes of 
OPAL for many valuable discussions, and the organisers and participants of 
TAU98 for such a productive and congenial workshop.
This work was supported in part by the National Science Foundation.

\end{document}